\documentclass[twocolumn,appendixfloats,tighten]{aastex6}
\usepackage{newtxtext,newtxmath}
\usepackage[T1]{fontenc}
\usepackage{ae,aecompl}

\usepackage{amsmath}	
\usepackage{amssymb}	

\usepackage{ctable}
\usepackage{url}
\usepackage{xspace}
\usepackage[normalem]{ulem}
\usepackage{hyperref}
\usepackage[all]{hypcap}
\usepackage{graphicx}

\def\msun{{\rm\,M_\odot}}

\def\msun{{\rm\,M_\odot}}

\newcommand{\kms}{\, {\rm km\, s}^{-1}}

\newcommand{\be}{\begin{equation}}
\newcommand{\ee}{\end{equation}}

\newcommand{\rsun}{ R_{\odot}}

\def\h2{${\rm\,H_2}$}

\usepackage{color}

\begin{document}

\title{Formation of mass gap objects in highly asymmetric mergers}

\author{Mohammadtaher Safarzadeh\altaffilmark{1}, Abraham Loeb\altaffilmark{1}}
\altaffiltext{1}{Center for Astrophysics | Harvard \& Smithsonian, 60 Garden Street, Cambridge, MA, USA
\href{mailto:msafarzadeh@cfa.harvard.edu}{msafarzadeh@cfa.harvard.edu}}

\begin{abstract}
The LIGO/Virgo Collaboration (LVC) recently reported the detection of GW190814, a merger of a $23^{+1.0}_{-0.9}~\msun$ primary black hole (BH), and a $2.6^{+0.08}_{-0.08}~\msun$ secondary.
The secondary's mass falls into the mass gap regime, which refers to the scarcity of compact objects in the mass range of 2-5 $\msun$.
The first clue to the formation of the GW190814 lies in the fact that the primary is a very massive BH. 
We suggest that the secondary was born as a neutron star (NS) where a significant amount of the supernova ejecta mass from its formation remained bound to the binary due to the presence of the massive BH companion.  
The bound mass forms a circumbinary accretion disk, and its accretion onto the NS created a mass gap object.
In this scenario, LIGO/Virgo will only detect mass gap objects in binary mergers with an extreme mass ratio. We also predict a correlation between the mass of the secondary and the mass of the primary in such asymmetric mergers.
Our model can be tested with future data from the LVC's third-observing run.

\end{abstract}

\section{Introduction}\label{sec:intro}

The LIGO/Virgo Collaboration (LVC) announced the detection of GW190814 \citep{Abbott:2020jo}, a compact binary merger with three unusual characteristics. 
First, the secondary compact object has a mass of about $2.6^{+0.08}_{-0.08}~\msun$ that lies in the mass gap regime. 
The mass gap term refers to the scarcity of black holes in the mass range 2-5 $\msun$ which is suggested to exist in nature based on the observations of black holes in low mass X-ray binaries \citep{Ozel:2010hd,Farr:2011ct}.
Regardless of whether this mass gap object is a neutron star (NS) or a BH, GW190814 is the first robust detection of an object in the mass gap range.

The second unusual fact about GW190814 is that its primary component is a massive BH with a mass of $23^{+1.0}_{-0.9}~\msun$. 
Such an extreme low mass ratio binary ($q=0.11$) would be difficult to account for in the main formation scenarios of binary compact objects. Broadly speaking, compact objects 
could be assembled either dynamically in dense stellar systems such as globular clusters \citep[e.g., ][]{Zwart:2004jj,Chatterjee:2016fl,Antonini:2017el,Rodriguez:2018ci,Gupta:2020kz}, 
or through binary stellar evolution in the field \citep[e.g., ][]{Flannery:1975to,Belczynski:2002gi,Dominik:2012cwa,Zaldarriaga:2017fn,Gerosa:2018hw,Bavera:2019ut}.
The dynamical assembly would not naturally yield a highly asymmetric mass ratio binary due to mass segregation in dense stellar environments. 
Formation of asymmetric binary black holes (BBHs) would also be suppressed in field binaries as processes such as mass transfer tend to equalize the mass of each compact object. 

Third, the merger rate of GW190814-type class of binaries is estimated to be between 1-23 $\rm year^{-1} Gpc^{-3}$. 
This high merger rate is only mildly less than the merger rate of BBHs with equal masses \citep{Abbottetal:2018vb}.

The mass gap object in GW190814 could be either formed by merging two NSs \citep[in fact, the remnant of GW170817 lies in this mass range; see ][]{Abbott:2017kt} or by a modification of the supernova engine model to allow the formation of objects in this mass range \citep{Fryer:2012jk}. The formation of a mass gap object alone is not enough to account for this system, and a successful model has to account for the merging of the mass gap object with a massive BH. 
Scenarios in which two NSs first merge to form a mass gap object, and the product itself merges with another BH are extremely rare in dense environments. 
An alternative scenario involving wide hierarchical quadruple systems was suggested by \citet{Safarzadeh:2020gs}, and variants of this scenario followed \citep[e.g., ][]{Fragione:2020dc}. 
However, none of these proposed channels can account for the high merger rate of the GW190814 class of binary mergers. 
Attempts to account for such mergers by modifying the supernova engine model still resulted in an order of magnitude lower merger rate than needed to account for GW190814, making this formation channel unlikely \citep{Zevin:2020vp}. 
While such objects can form through accretion in AGN disks, their formation rate is highly uncertain \citep{Yang2020}.
Moreover, \citet{Vattis:2020ub} show that a primordial BH origin for the mass gap object in GW190814 is unlikely.

In this \emph{Letter}, we suggest an alternative channel capable of reproducing GW190814-type mergers.
We show that it is possible to retain a large fraction of supernova mass ejecta bound to the binary due to the presence of a massive BH in the system. 
In this scenario, a supernova (SN) explosion first forms a NS, and a large fraction of the ejecta mass remains bound to the binary and forms a circumbinary accretion disk. 
The NS will accrete from the accretion disk and transition into becoming a mass gap object.  

The structure of this \emph{Letter} is as follows. In \S2 we describe the requirement for ejecta mass trapping in a binary system. 
In \S3 we discuss the caveats of our model, and in \S4 we summarize our results and main conclusions.

\section{Bound ejecta mass}

Gravitational-wave emission from a compact binary object leads to a merger of the binary's components on a timescale \citep{PetersPC:1964bc},

\be
t=\frac{5}{256} \frac{c^5}{G^3}\frac{a^4}{(M_1 M_2)(M_1+M_2)},
\ee
where $M_1$, and $M_2$ are the mass of the primary and secondary compact objects, and $a$ is the semi-major axis.
A binary with characteristics similar to GW190814 will merge within a Hubble time if its initial separation is less than about 20 $\rsun$ \citep{Jani:2020hl} in the absence of eccentricity. 
Population synthesis models suggest two distinct channels that can lead to the formation of GW190814 \citep{Zevin:2020vp}.
In the first channel (labeled A), two phases of mass transfer take place and after $\sim10$ Myr the binary consists of a 21.6 $\msun$ BH in orbit around a
7.7 $\msun$ He star with orbital separation of 1.1 AU. Following the SN explosion of the He star, a 3 $\msun$ compact object is formed.
This event results in a highly eccentric orbit with separation $a=0.65$ AU and eccentricity of $e=0.99$, which merges in 14 Myrs. 
This particular example was chosen to illustrate a system that merges within a Hubble time, while binaries that don't get into an eccentric orbit are common.
We note that \citet{Zevin:2020vp} adopt a delayed SN prescription to be able to form mass gap objects.
In their second formation channel (labeled B), it is the mass gap object that forms first, and at $t=8.3$ Myr the system
consists of a 3 $\msun$ mass gap object in orbit around a 23.4 $\msun$ He star at separation $a=0.02$ AU, leading to a merger within about $\sim4$ Myrs. 
It is the channel A that dominates the formation of GW190814-like systems. 
However, since both formation channels lead to a merger within tens of Myr, we adopt a maximum separation between the progenitor He star and the primary BH 
to be less than about $\sim4~\rsun$ such that the binary merges on a similar timescale. 

We note that the examples presented in \citet{Zevin:2020vp}, which merge within tens of Myrs do not represent a typical merging timescale of their GW190814-like binaries. 
In fact, most of their GW190814-like binaries have extremely large separations and only merge within a Hubble time due to their eccentricity. 
Since we do not incorporate eccentricity, our binaries have to be on a tighter orbit to merge within a Hubble time. If the predictions of our model come true in light of the LIGO/Virgo third-observing run data, 
this work could guide population synthesis models to adjust their parameters to account for this model.

We assume that the secondary is born as a NS. The progenitor of the NS is a He star, as its Hydrogen envelope is stripped through a mass transfer phase.
The NS mass depends on the pre-SN mass of the He star. 
A detailed study of the relation between the remnant mass of the NSs from the explosion of He stars was performed by \citet{Ertl:2020ga}.
NSs with a gravitational mass of about 1.4 $\msun$ have pre-SN progenitors with a mass between 3-7 $\msun$ (see their Figure 15, top panel). A pre-SN mass in that range are the outcome of He stars with an initial mass below $\leq10~\msun$.
Moreover, the explosive energy for such pre-SN masses ranges between 0.1-1 $\times10^{51}\rm erg~s^{-1}$ (see the top-left panel in their Figure 14).

A fraction of the ejecta material (the difference between the pre-SN mass and the final remnant mass) in the final SN explosion leading to the formation of a NS can remain bound to the binary. 
This is a result of the presence of a massive BH companion in orbit around the pre-SN progenitor star. 
The fraction of bound ejecta mass depends on several parameters. First, the more massive the companion is, the larger is the fraction of the supernova ejecta that will remain bound to the system. 
Moreover, the separation between the massive BH and the He star, the density and velocity profile of the ejecta, and the explosion energy of the SN are a few other parameters that influence the bound mass fraction. 
If enough material remains bound to the binary, this material can get accreted onto the newly born NS increasing its mass into the mass gap regime. 
If this is the main formation channel for GW190814-like sources, we expect to find mass gap objects mainly in the company of very massive BHs.
This channel would predict that the mass of the mass gap object should correlate with the mass of its BH companion, transitioning from a NS mass range to a mass gap BH with increasing mass of its companion BH.

\begin{figure*}
\includegraphics[width=1\columnwidth]{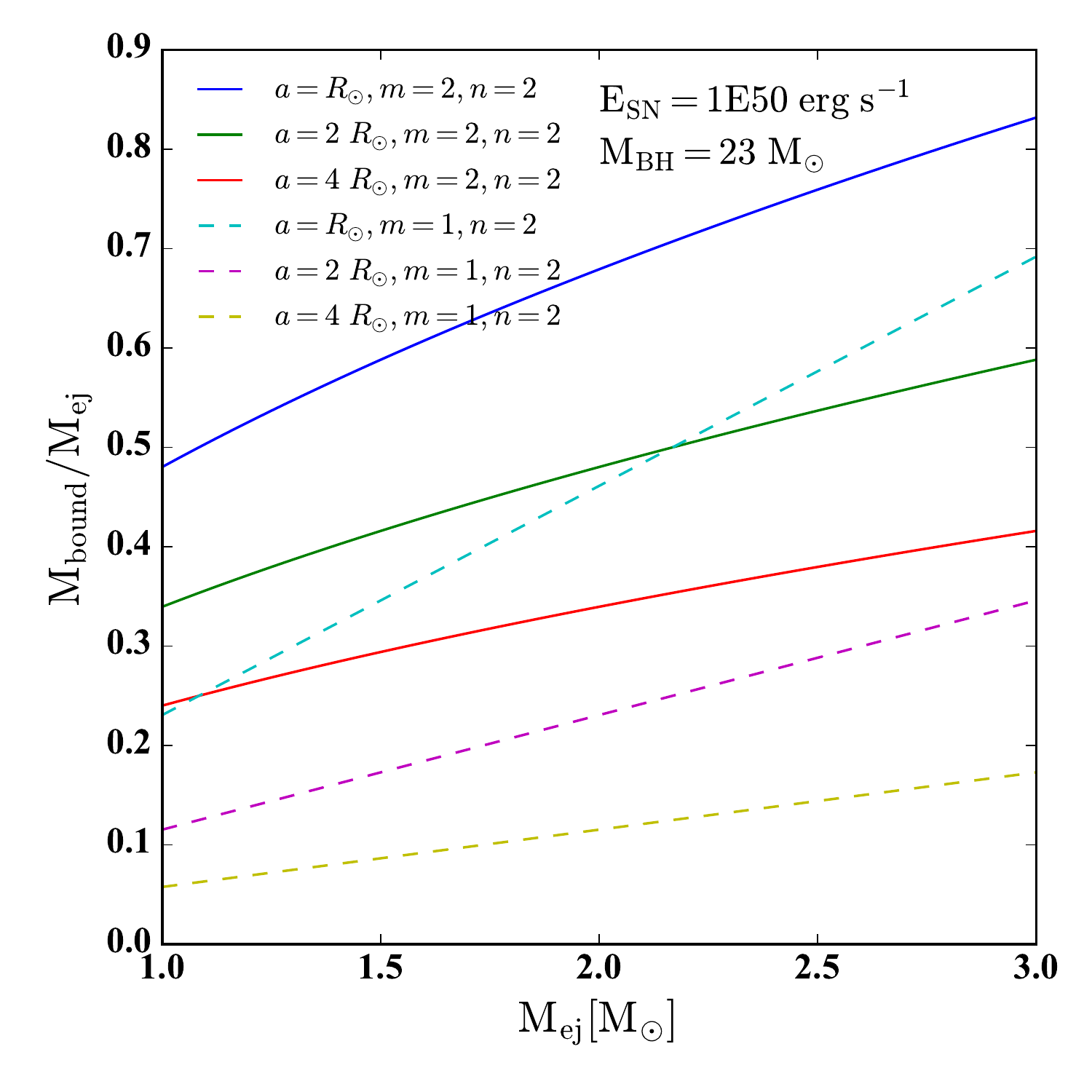}
\includegraphics[width=1\columnwidth]{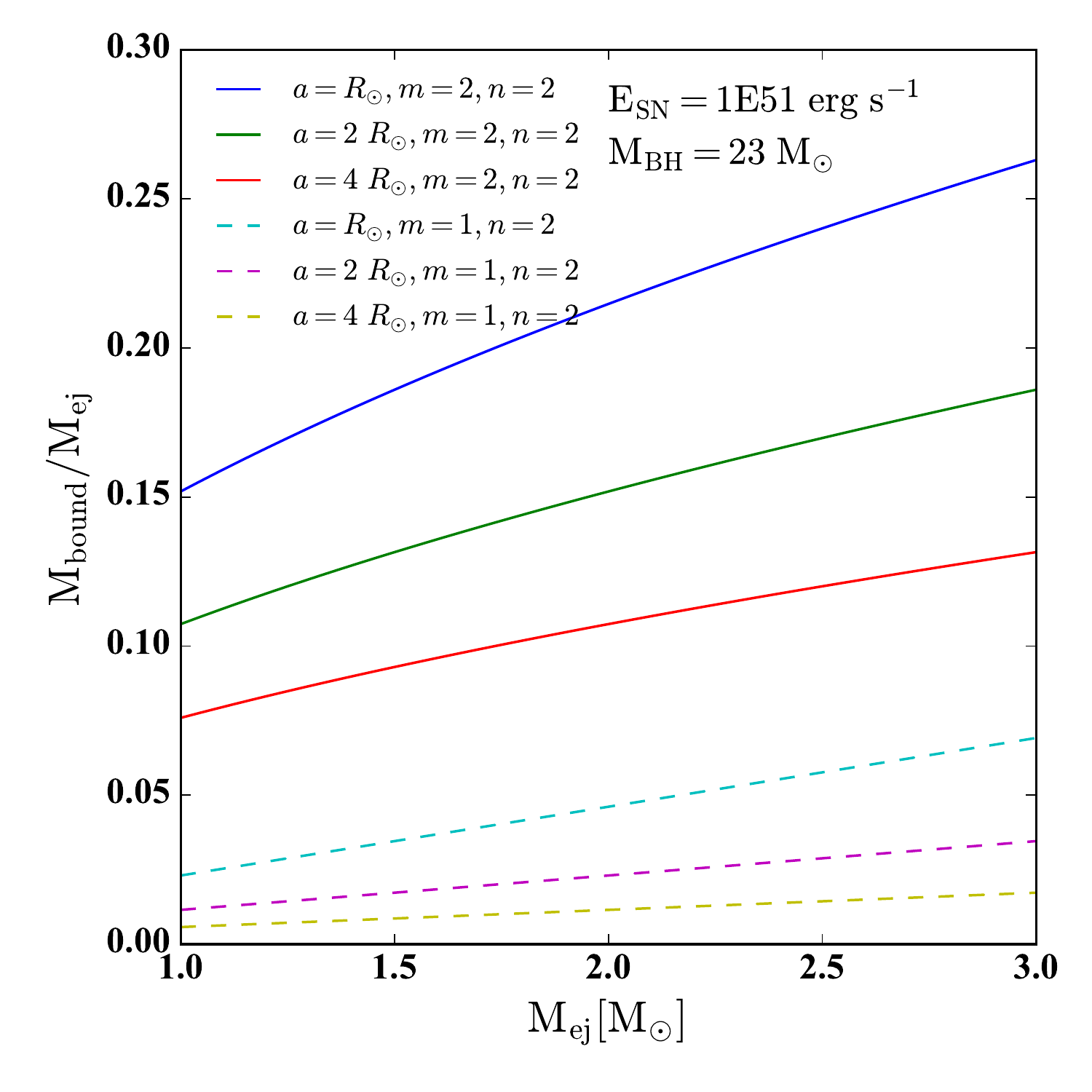}
\caption{\emph{Left panel:} The fraction of the ejecta mass that remains bound to the binary when the secondary's SN explosion takes place with a mass $M_{\rm ej}$ of ejecta material. 
The bound mass depends on the mass of the BH companion and its distance from the secondary when the SN explosion takes place. We set the companion BH mass to be $23~\msun$ and show the dependence of the bound mass on the orbital separation. 
We assume a maximum orbital separation for the binary to be about $4~\rsun$, requiring the binary to merge within tens of Myrs consistent with population synthesis results of \citet{Zevin:2020vp}.
The bound mass would also depend on the density profile of the ejecta. The solid lines refer to a profile with no change in density slope, and the dashed lines refer to a profile that transition from a shallow density profile with a slope of $m=2$ to a steep profile with a slope $n=3$. We assume the explosion energy to be $10^{50}\rm erg~s^{-1}$ \citep{Ertl:2020ga}. For example, a solar mass of bound material requires the total ejecta material to be $\approx1.5~\msun$ if the separation between the pre-SN He star and the $23~\msun$ is about $a=\rsun$ (blue line). \emph{Right panel:} The same, but for the case where $E_{SN}=10^{51}\rm erg~s^{-1}$.
}
\label{fig:1}
\end{figure*}

\begin{figure*}
\includegraphics[width=1\columnwidth]{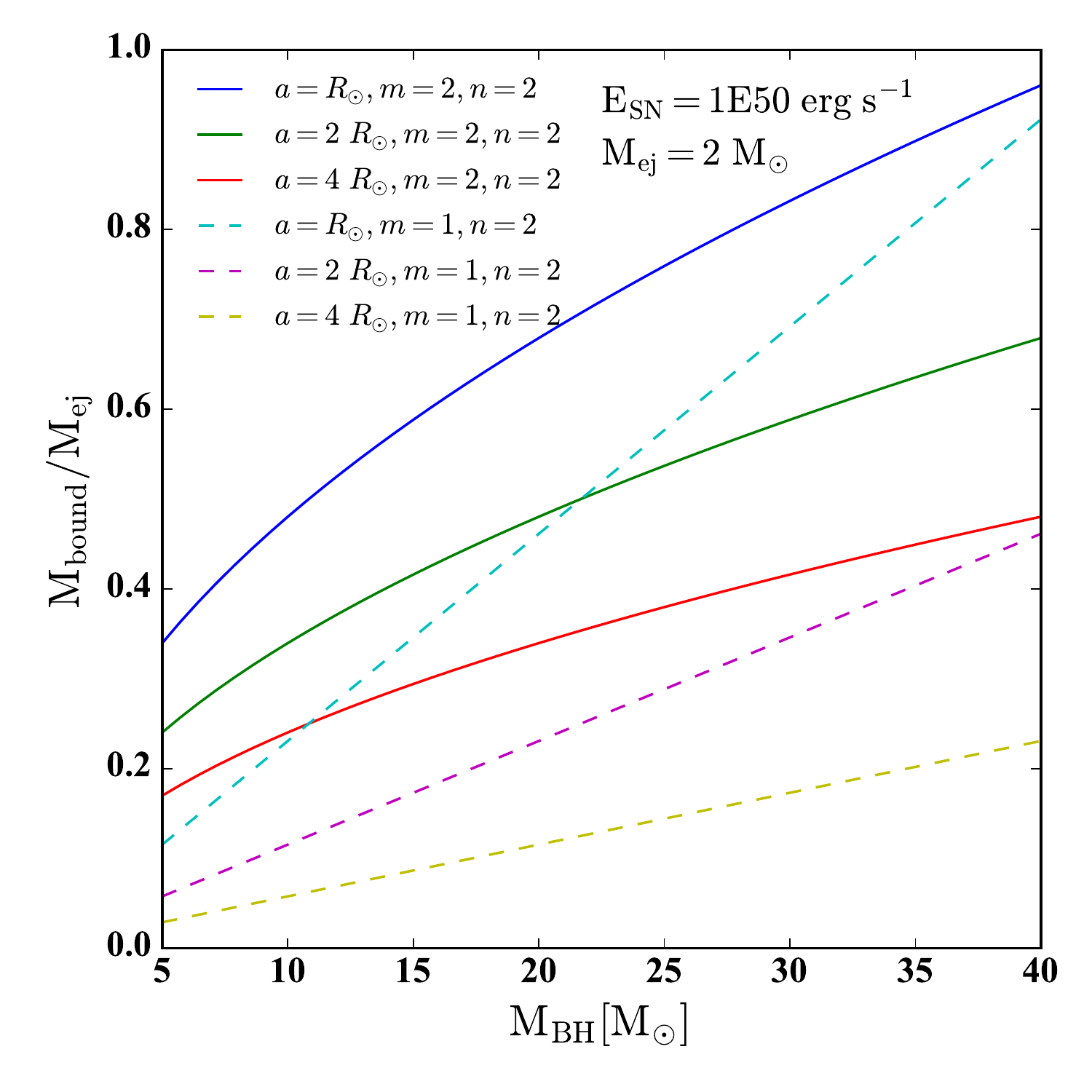}
\includegraphics[width=1\columnwidth]{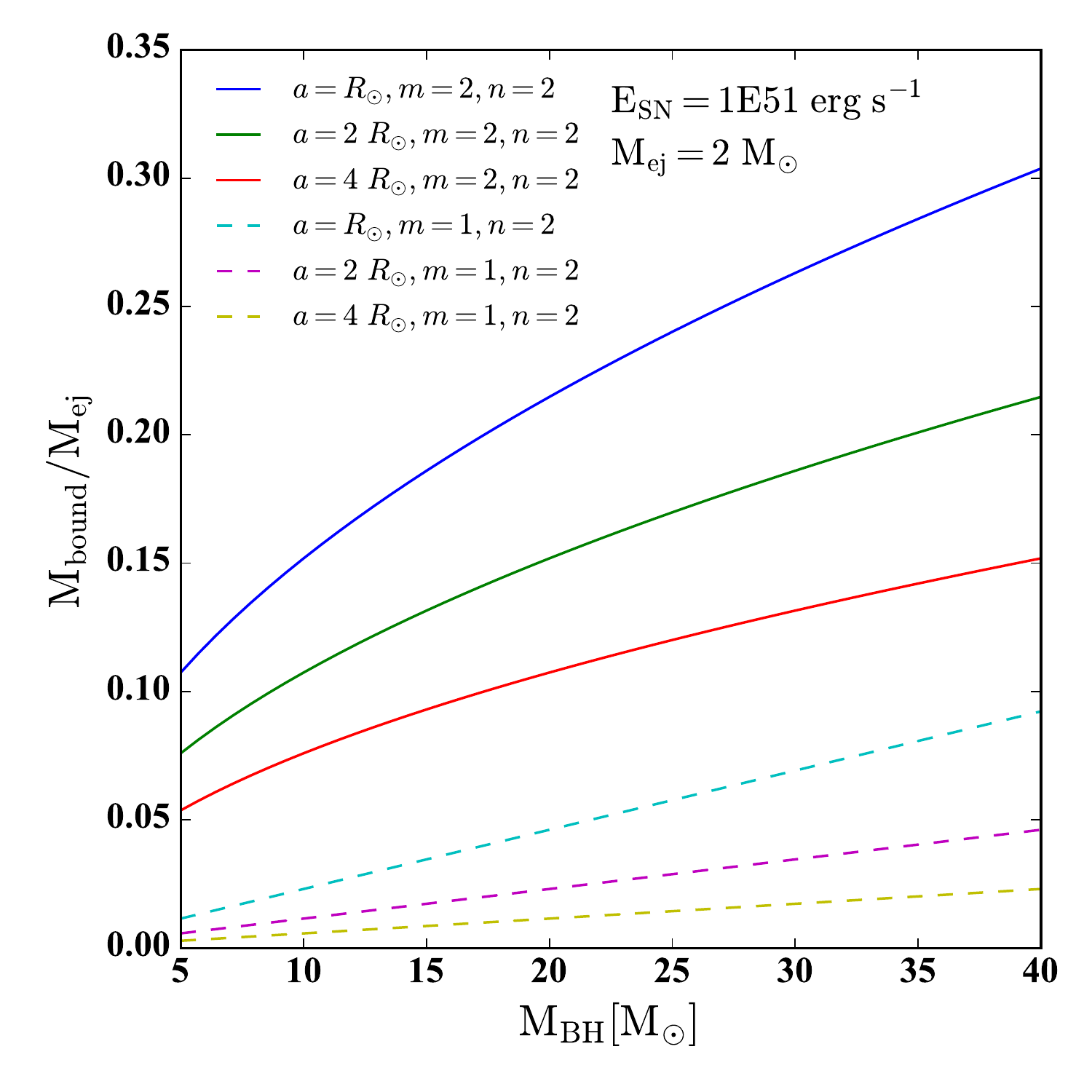}
\caption{\emph{Left panel:} The fraction of the ejecta mass that remains bound to the binary when the secondary's SN explosion takes place as a function of the companion BH mass. 
We assume during the SN explosion of the secondary's progenitor, $2~\msun$ of material is ejected, and $E_{SN}=10^{50}\rm erg~s^{-1}$. 
The bound fraction depends on many parameters; however, it is possible for half of the ejecta to remain bound to the binary if the companion is a 20 $\msun$ BH with a separation $a\approx\rsun$. 
The result depends on the assumed density profile of the ejecta, but this dependence becomes less important at smaller orbital separations. \emph{Right panel:} The same but for $E_{SN}=10^{51}\rm erg~s^{-1}$.
}
\label{fig:2}
\end{figure*}

We follow the formalism presented in \citet{Suzuki:2016fz} for the supernova mass ejecta density and velocity profile, which is based on the results of \citet{Truelove:1999it}.
The ejecta is assumed to expand in a spherical and homologous fashion, with the velocity profile at time $t$ given by,
\begin{equation}
v(t,R)=\left\{\begin{array}{ccl}
R/t&\mathrm{for}&R\leq v_\mathrm{ej}t,\\
0&\mathrm{for}&v_\mathrm{ej}t<R,
\end{array}\right.
\label{eq:ejecta_velocity}
\end{equation}
where $v_\mathrm{ej}$ denotes the maximum velocity of the ejecta. 
The ejecta is assumed to be composed of two components, one with a shallow and the other with a steep density gradient for the inner, and outer regions respectively. 
The density structures are modeled in terms of a power-law function of the velocity, $\rho\propto v^{-m}$ for the inner and $\rho\propto v^{-n}$ for the outer ejecta. 
The mass density profile is described as follows,
\begin{equation}
\rho(t,R)=\left\{\begin{array}{ccl}
\frac{f_\mathrm{3}M_\mathrm{ej}}{4\pi \omega_\mathrm{c}^3v_\mathrm{ej}^3t^3}\left(\frac{R}{w_\mathrm{c}v_\mathrm{ej}t}\right)^{-m}&\mathrm{, for}&R\leq w_\mathrm{c}v_\mathrm{ej}t,\\
\frac{f_\mathrm{3}M_\mathrm{ej}}{4\pi \omega_\mathrm{c}^3v_\mathrm{ej}^3t^3}\left(\frac{R}{w_\mathrm{c}v_\mathrm{ej}t}\right)^{-n}&\mathrm{, for}&w_\mathrm{c}v_\mathrm{ej}t<R\leq v_\mathrm{ej}t,\\
0&\mathrm{, for}&v_\mathrm{ej}t<R,
\end{array}\right.
\label{eq:ejecta_density}
\end{equation}
with a numerical factor $f_{l}$ given by,
\begin{equation}
f_l=\frac{(n-l)(l-m)}{n-m-(l-m)w_\mathrm{c}^{n-l}}.
\end{equation}
The integration of $4\pi R^2\rho(t,R)$ over the radius $R$ from $0$ to $vt$ gives the mass $M(v)$ of the ejecta travelling at velocities slower than $v$,
\begin{eqnarray}
M(<v)&=&\int_0^{vt}4\pi R^2\rho(t,R)dR
\nonumber\\
&=&
\left\{\begin{array}{llc}
\frac{f_\mathrm{3}M_\mathrm{ej}}{3-m}\left(\frac{v}{w_\mathrm{c}v_\mathrm{ej}}\right)^{3-m}
&\mathrm{, for}&v\leq w_\mathrm{c}v_\mathrm{ej},\\
\frac{f_\mathrm{3}M_\mathrm{ej}}{3-n}
\left[\left(\frac{v}{w_\mathrm{c}v_\mathrm{ej}}\right)^{3-n}
-\frac{n-m}{3-m}
\right]
&\mathrm{, for}&w_\mathrm{c}v_\mathrm{ej}<v.\\
\end{array}\right.
\end{eqnarray}
The parameter $w_\mathrm{c}$ indicates the location of the interface between the inner and outer ejecta in the velocity coordinate which we set to a value $=1$.

When the supernova explosion takes place, the ejecta with a velocity less than the escape velocity of the system will remain bound to the binary and form a circumbinary accretion disk.
The escape velocity of the ejecta depends on the companion BH mass and semi-major axis of the binary,
\be
V_{\rm esc}=\left(\frac{2GM_{BH}}{a}\right)^{1/2}.
\ee
For GW190814, we assume a maximum binary separation of $4~\rsun$, and a BH mass is about $23~\msun$. 

Figure \ref{fig:1} shows the fraction of the total ejecta mass will remain bound to the binary in the presence of the massive BH. 
We assume channel A in \citet{Zevin:2020vp} in which a massive BH is already formed, and the secondary progenitor makes a NS through a SN explosion.
The solid lines in the left panel of Figure \ref{fig:1} show the amount of bound mass at a given ejecta mass, assuming a constant density profile with slopes $m=n=2$. 
Each solid line refers to a specific separation between the BH and the pre-SN object. 
For example, if the SN explosion takes place when the BH and the pre-SN object have a solar radii separation, about 1 $\msun$ of material will remain bound to the binary if the ejecta mass is $\approx1.5~\msun$ (blue line). 
The dashed lines show the same results but assuming a shallower inner density profile for the ejecta. 
The right panel shows the same, but assuming the explosion energy of $E_{\rm SN}=10^{51} \rm erg~s^{-1}$.

We explicitly explore the role of companion BH mass on the bound fraction of the ejecta material. 
The left panel of figure \ref{fig:2} shows the results assuming 2 $\msun$ of ejecta material with an explosion energy of $10^{50} \rm erg~s^{-1}$ and different 
separations between the pre-SN He star and the BH. 
About half of the ejecta material can remain bound to the binary if the companion BH has a mass of 20 $\msun$ with separations in the range 1-2 $\rsun$. The bound mass depends on many parameters. 
For example, the ejecta profile plays an important role, though decreasing in its significance for small orbital separations. Moreover, large explosion energy, as shown in the right panel of Figure \ref{fig:2}, could reduce the bound fraction.

Can the NS accrete about 1 $\msun$ of the bound ejecta material before it merges with its massive companion BH? The Eddington accretion rate limit onto a NS is $\dot{M}_{\rm Edd}=2.8\times10^{17}\rm g~s^{-1}$.
Assuming the binary merges within tens of Myrs after the formation of the NS, the NS needs to accrete the 1 $\msun$ at ten times the Eddington accretion rate limit. 
Such accretion rates are inferred around NS detected as ultra-luminous X-ray source \citep{Kaaret:2017eb}. 

\begin{figure*}
\includegraphics[width=1\columnwidth]{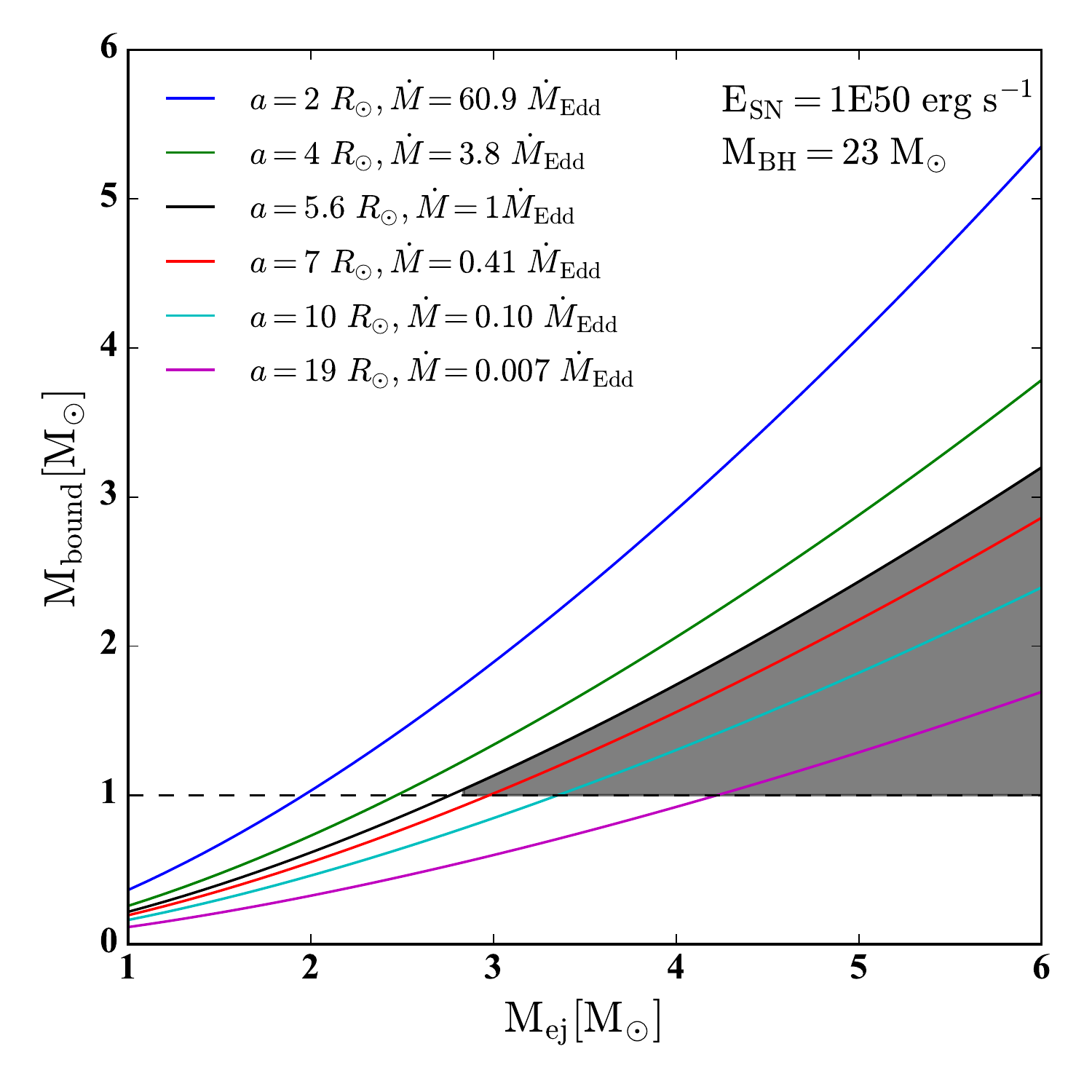}
\includegraphics[width=1\columnwidth]{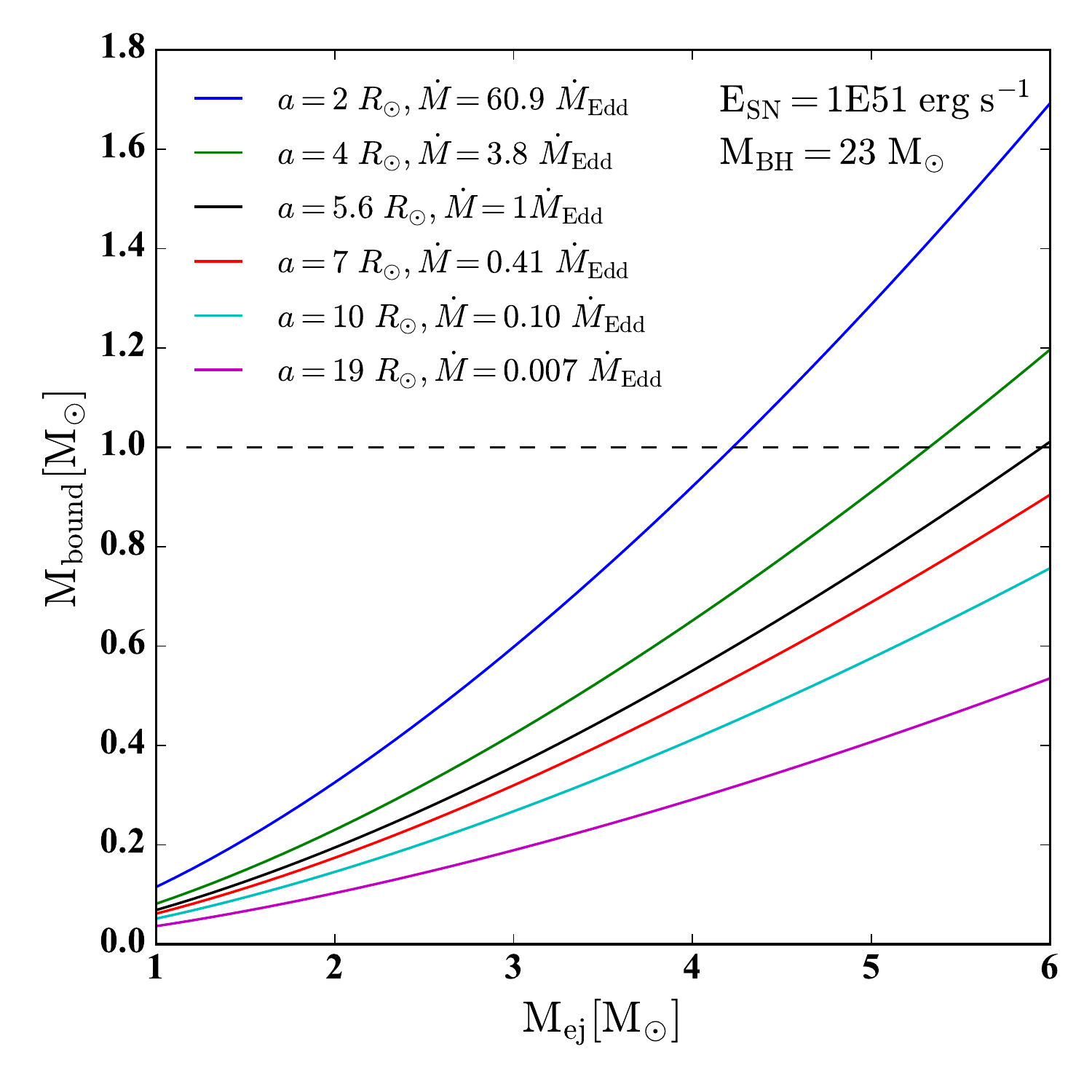}
\caption{\emph{Left panel:} The bound mass as a function of the ejecta mass. Each line indicates an initial orbital separation between the pre-SN progenitor of the NS and the 23 $\msun$ BH.
The accretion rate required to accrete 1 $\msun$ of bound material onto a newly born NS for each separation is indicated in the legend in units of Eddington accretion rate for a 1.4 $\msun$ NS ($\dot{M}_{\rm Edd}=2.8\times10^{17}\rm g~s^{-1}$).
We assume an explosion energy of $E_{\rm SN}=10^{50} \rm erg~s^{-1}$.  
The larger the initial separation, the longer the merger time, and therefore a lower accretion rate is required for the NS to be able to accrete 1 $\msun$ of the bound material to transition into a mass gap object.
The shaded region shows the parameter space where accretion rates lower than the Eddington limit would be sufficient. 
 \emph{Right panel:} The same but for the case of $E_{SN}=10^{51}\rm erg~s^{-1}$. In this case, a mass gap object forms only through a super Eddington accretion rate.
}
\label{fig:3}
\end{figure*}

Figure \ref{fig:3} shows the accretion rate required to accrete 1 $\msun$ of the bound material onto a newly born NS as a function of the ejecta mass. 
We assume the companion BH has a mass of 23 $\msun$. We assume different orbital separation for the binary at the onset of the SN explosion. At larger separations, a smaller ejecta mass remains bound to the binary. However, the merger time increases $(t\propto a^4)$ leaving more time for the NS to accrete 1 $\msun$ of material. 
This, in turn, leads to requiring lower accretion rates. Each line in Figure \ref{fig:3} assumes a specific orbital separation with a maximum value set to 19 $\rsun$ such that the merger timescale does not exceed the Hubble time.
For each orbital separation, the accretion rate in terms of the Eddington limit is indicated in the legend. 
The left panel shows the results for an explosion energy of $E_{\rm SN}=10^{50} \rm erg~s^{-1}$, and the right panel assumes $E_{\rm SN}=10^{51} \rm erg~s^{-1}$. 

Figure \ref{fig:3} shows that the required accretion rate exceeds the Eddington limit for orbital separations smaller than about $5.6~\rsun$ (black line). 
The shaded region in the left panel of Figure \ref{fig:3} is the parameter space in which an accretion rate less than the Eddington limit is enough to transform the NS into a mass gap object by accreting 1 $\msun$ of the bound material. 
The right panel shows the result for $E_{\rm SN}=10^{51} \rm erg~s^{-1}$, where a super Eddington accretion rate is required to form a mass gap object. We note that for all panels, we have assumed a velocity profile with $m=n=2$.

\section{Two caveats}
We note that the bound mass calculated in the previous section is subject to two caveats: 
({\it i}) the orbital velocity of the pre-SN compact object, which in the case of small separations is non-negligible, should be taken into account; and ({\it ii})
the above method assumes the ejecta material is directed radially outward from the companion BH; 
however, a larger fraction of the ejecta material that travels towards the BH remains bound to the system, and similarly a lower fraction of the ejecta that 
travels along the orbital velocity vector will remain bound to the system. 

To account for these effects, we compute the boost due to the orbital velocity to the ejecta material at each solid angle based on the magnitude of the vectorial sum of $\vec{v_{\rm ej}}$, and $\vec{v_{\rm orb}}$ at each solid angle.
At each solid angle, we compute the ejecta velocity at which the magnitude of the sum of the two vectors exceeds the escape velocity of the system. Then we perform the integral over all solid angles:
\be
M_{\rm bound}=\int M(v<v_{\rm esc})d\Omega.
\ee
in order to get the total amount of bound material due to the effect of orbital velocity. We find $M_{\rm bound}\approx 0.8 M_{\rm bound}^{w/o}$, 
where $M_{\rm bound}^{w/o}$ stands for the total bound material without taking into account the impact of the orbital velocity vector. Therefore, a correction of 20\% needs to be applied to the reported bound mass fractions. 

Another factor is that a NS receives a natal kick at birth, which can potentially unbind the system. In that case, the newly born NS leaves, and there is no merger with the 23 $\msun$ BH. 
If the binary survives the NS natal kick, the orbit becomes eccentric. Even in the absence of such kicks, the ejected mass will make the orbit eccentric. 
However, since the orbital speed of the pre-SN compact object is of order $10^3 ~\kms$, and natal kicks are of order $10^2 ~\kms$, we expect the resulting eccentricity to play a minor role in the subsequent evolution of the system. 
Moreover, \citet{Suwa:2015bs} showed that the natal kicks of NSs born from stripped stars are smaller than those NSs born from single star progenitors.

Hydrodynamical simulations of such a process, although in a rather different context, have been performed before \citep[e.g., ][]{Rimoldi:2015up,Schroder:2018ko}. 
The order of magnitude estimates that we present here should be refined through hydrodynamical simulations in the future. 

The bound material forms a disk around the binary from where it gets accreted to either the NS or the BH. 
However, most of this mass is expected to be accreted onto the NS as the massive BH stays near the binary's center of mass, while the NS sweeps through the accretion disk. 
Such a system resembles a compact binary inside an accretion disk around a supermassive BH at the center of a galaxy \citep[e.g., ][]{Tagawa:2019tu}.
Simulation of circumbinary accretion disk around binary black holes have been performed in the literature, and the effect of the binary's mass ratio has been explored \citep[e.g., ][]{DOrazio:2016gy,Duffell:2019vu}.
In particular, \citet{Duffell:2019vu} find the following fitting formula for the accretion rate onto the components of the binary:
\be
\frac{\dot{m_2}}{\dot{m_1}}=\frac{1}{0.1+0.9q},
\ee
 where $m_1$, and $m_2$ are the primary, and the secondary component masses, and $q=m_2/m_1$ is the mass ratio of the system. 
 Given the parameters of our system ($m_1=23~\msun$, $m_2=1.4\msun$) more than 90\% of the gas is going to be accreted onto the neutron star.
 The circumbinary accretion disk simulations assume that the disk mass is negligible in the dynamics of the system. The Toomre Q parameter \citep{Toomre:1964fe} quantifies the stability of accretion disk against gravitational fragmentation:
 \be
 Q=\frac{C_s\Omega}{\pi G \Sigma}=\frac{M_{BH}}{M_d}\frac{H}{r},
 \ee
  where $C_s$, $\Omega$, and $\Sigma$ are the sound speed, orbital frequency, and surface density of the disk. $H/r$ is the disk height to radius ratio, which for a thin accretion disk is below 0.1 (see \citet{Kratter:2016dw} for a review), 
  and $M_d$ is the disk mass. 
  Given the parameters of our system, disk masses below about $2.5\msun$ are going to be stable against fragmentation by self-gravity. 
  Radiation pressure further supports disk against fragmentation. In particular we expect the neutron star to heat up the disk, permitting larger disk masses. 
  Numerical simulations would be needed to refine our predictions.
 
\section{Conclusions}
It is challenging to understand the formation pathway of GW190814. Previous models fail to account for either the high merger rate inferred from this system, the presence of a mass gap object, or the binary's highly asymmetric masses. 
We have shown that the capture of the SN ejecta material in the presence of a massive BH companion holds the key to understanding the formation of this system. 
Whereas without a companion the ejecta material leaves the system, the presence of a massive BH deepens the gravitational potential well of the system so that a non-negligible fraction of the ejecta material remains bound to the binary. 
This material later on gets accreted to the NS formed from the explosion of the secondary's progenitor He star, increasing its mass into the mass gap range. 

This process depends on the explosion energy of the pre-SN He stars, the ejecta mass, the orbital separation between the BH and the pre-SN He star, the mass of the companion BH, and the SN ejecta density and velocity profile.
However, we have shown that if the BH companion is massive, there exists a parameter space in which the bound fraction becomes comparable to the amount of mass needed for a NS to become a mass gap object.  

If our model holds, we predict mass gap events detected by the LVC will be similar to GW190814, in that a mass gap object is accompanied with a much more massive BH. 
This would be in contrast with models in which mass gap objects form from SNe fallback \citep{Ertl:2020ga}. 
For example, \citet{Sukhbold:2018br} show that massive ($\approx 2~\msun$) neutron stars can form from stars with Zero Age Main Sequence (ZAMS) masses between 12 and 15 $\msun$. Therefore, small amount of fallback material would permit the formation of mass gap objects. 
Unlike our circumbinary accretion model, a fallback would not require a massive binary companion for the formation of a mass gap object, a key difference that could be tested with future LVC data.
Moreover, if the secondary is a highly spinning neutron star \citep{Most:2020uu}, its spin angular momentum could be supplied through the circumbinary accretion disk in our model.
We also predict that the mass of the mass gap object would be correlated with its BH companion mass, an expectation that could be tested with the upcoming release of the third observing run of LVC.

\acknowledgements 
We thank the referee for insightful comments, and Zoltan Haiman, Mike Zevin, and Enrico Ramirez-Ruiz for comments on the earlier version of this manuscript. 
This work is supported by the National Science Foundation under Grant No. AST-1440254, and by Harvard's Black Hole Initiative, which is funded by JTF and GBMF. 

\bibliographystyle{apj}
\bibliography{the_entire_lib.bib}

\begin{thebibliography}{}
\expandafter\ifx\csname natexlab\endcsname\relax\def\natexlab#1{#1}\fi

\bibitem[{Abbott {et~al.}(2017)Abbott, Abbott, Abbott, Acernese, Ackley, Adams,
  Adams, Addesso, Adhikari, Adya, Affeldt, Afrough, Agarwal, Agathos, Agatsuma,
  Aggarwal, Aguiar, Aiello, Ain, Ajith, Allen, Allen, Allocca, Altin, Amato,
  Ananyeva, Anderson, Anderson, Angelova, Antier, Appert, Arai, Araya, Areeda,
  Arnaud, Arun, Ascenzi, Ashton, Ast, Aston, Astone, Atallah, Aufmuth, Aulbert,
  AultONeal, Austin, Avila-Alvarez, Babak, Bacon, Bader, Bae, Bailes, Baker,
  Baldaccini, Ballardin, Ballmer, Banagiri, Barayoga, Barclay, Barish, Barker,
  Barkett, Barone, Barr, Barsotti, Barsuglia, Barta, Barthelmy, Bartlett,
  Bartos, Bassiri, Basti, Batch, Bawaj, Bayley, Bazzan, B{\'e}csy, Beer,
  Bejger, Belahcene, Bell, Berger, Bergmann, Bernuzzi, Bero, Berry, Bersanetti,
  Bertolini, Betzwieser, Bhagwat, Bhandare, Bilenko, Billingsley, Billman,
  Birch, Birney, Birnholtz, Biscans, Biscoveanu, Bisht, Bitossi, Biwer,
  Bizouard, Blackburn, Blackman, Blair, Blair, Blair, Bloemen, Bock, Bode,
  Boer, Bogaert, Bohe, Bondu, Bonilla, Bonnand, Boom, Bork, Boschi, Bose,
  Bossie, Bouffanais, Bozzi, Bradaschia, Brady, Branchesi, Brau, Briant,
  Brillet, Brinkmann, Brisson, Brockill, Broida, Brooks, Brown, Brown, Brunett,
  Buchanan, Buikema, Bulik, Bulten, Buonanno, Buskulic, Buy, Byer, Cabero,
  Cadonati, Cagnoli, Cahillane, Calder{\'o}n~Bustillo, Callister, Calloni,
  Camp, Canepa, Canizares, Cannon, Cao, Cao, Capano, Capocasa, Carbognani,
  Caride, Carney, Carullo, Casanueva~Diaz, Casentini, Caudill, Cavagli{\`a},
  Cavalier, Cavalieri, Cella, Cepeda, Cerd{\'a}-Dur{\'a}n, Cerretani, Cesarini,
  Chamberlin, Chan, Chao, Charlton, Chase, Chassande-Mottin, Chatterjee,
  Chatziioannou, Cheeseboro, Chen, Chen, Chen, Cheng, Chia, Chincarini,
  Chiummo, Chmiel, Cho, Cho, Chow, Christensen, Chu, Chua, Chua, Chung, Chung,
  Ciani, Ciolfi, Cirelli, Cirone, Clara, Clark, Clearwater, Cleva, Cocchieri,
  Coccia, Cohadon, Cohen, Colla, Collette, Cominsky, Constancio, Conti, Cooper,
  Corban, Corbitt, Cordero-Carri{\'o}n, Corley, Cornish, Corsi, Cortese, Costa,
  Coughlin, Coughlin, Coulon, Countryman, Couvares, Covas, Cowan, Coward,
  Cowart, Coyne, Coyne, Creighton, Creighton, Cripe, Crowder, Cullen, Cumming,
  Cunningham, Cuoco, Dal~Canton, D{\'a}lya, Danilishin, D'Antonio, Danzmann,
  Dasgupta, Da~Silva~Costa, Dattilo, Dave, Davier, Davis, Daw, Day, De, DeBra,
  Degallaix, De~Laurentis, Del{\'e}glise, Del~Pozzo, Demos, Denker, Dent,
  De~Pietri, Dergachev, De~Rosa, DeRosa, De~Rossi, DeSalvo, de~Varona,
  Devenson, Dhurandhar, D{\'\i}az, Dietrich, Di~Fiore, Di~Giovanni,
  Di~Girolamo, Di~Lieto, Di~Pace, Di~Palma, Di~Renzo, Doctor, Dolique, Donovan,
  Dooley, Doravari, Dorrington, Douglas, Dovale~{\'A}lvarez, Downes, Drago,
  Dreissigacker, Driggers, Du, Ducrot, Dudi, Dupej, Dwyer, Edo, Edwards, \&
  Ef...}]{Abbott:2017kt}
Abbott, B.~P., Abbott, R., Abbott, T.~D., {et~al.} 2017, Physical Review
  Letters, 119, 161101

\bibitem[{Abbott {et~al.}(2020)Abbott, Abbott, Abraham, Acernese, Ackley,
  Adams, Adhikari, Adya, Affeldt, Agathos, Agatsuma, Aggarwal, Aguiar, Aich,
  Aiello, Ain, Ajith, Akcay, Allen, Allocca, Altin, Amato, Anand, Ananyeva,
  Anderson, Anderson, Angelova, Ansoldi, Antier, Appert, Arai, Araya, Areeda,
  Ar{\`e}ne, Arnaud, Aronson, Arun, Asali, Ascenzi, Ashton, Aston, Astone,
  Aubin, Aufmuth, AultONeal, Austin, Avendano, Babak, Bacon, Badaracco, Bader,
  Bae, Baer, Baird, Baldaccini, Ballardin, Ballmer, Bals, Balsamo, Baltus,
  Banagiri, Bankar, Bankar, Barayoga, Barbieri, Barish, Barker, Barkett,
  Barneo, Barone, Barr, Barsotti, Barsuglia, Barta, Bartlett, Bartos, Bassiri,
  Basti, Bawaj, Bayley, Bazzan, B{\'e}csy, Bejger, Belahcene, Bell, Beniwal,
  Benjamin, Benkel, Bentley, Bergamin, Berger, Bergmann, Bernuzzi, Berry,
  Bersanetti, Bertolini, Betzwieser, Bhandare, Bhandari, Bidler, Biggs,
  Bilenko, Billingsley, Birney, Birnholtz, Biscans, Bischi, Biscoveanu, Bisht,
  Bissenbayeva, Bitossi, Bizouard, Blackburn, Blackman, Blair, Blair, Blair,
  Bobba, Bode, Boer, Boetzel, Bogaert, Bondu, Bonilla, Bonnand, Booker, Boom,
  Bork, Boschi, Bose, Bossilkov, Bosveld, Bouffanais, Bozzi, Bradaschia, Brady,
  Bramley, Branchesi, Brau, Breschi, Briant, Briggs, Brighenti, Brillet,
  Brinkmann, Brito, Brockill, Brooks, Brooks, Brown, Brunett, Bruno, Bruntz,
  Buikema, Bulik, Bulten, Buonanno, Buskulic, Byer, Cabero, Cadonati, Cagnoli,
  Cahillane, Bustillo, Callaghan, Callister, Calloni, Camp, Canepa, Cannon,
  Cao, Cao, Carapella, Carbognani, Caride, Carney, Carullo, Diaz, Casentini,
  Casta{\~n}eda, Caudill, Cavagli{\`a}, Cavalier, Cavalieri, Cella,
  Cerd{\'a}-Dur{\'a}n, Cesarini, Chaibi, Chakravarti, Chan, Chan, Chao,
  Charlton, Chase, Chassande-Mottin, Chatterjee, Chaturvedi, Chatziioannou,
  Chen, Chen, Chen, Cheng, Cheong, Chia, Chiadini, Chierici, Chincarini,
  Chiummo, Cho, Cho, Cho, Christensen, Chu, Chua, Chung, Chung, Ciani,
  Ciecielag, Cie{\'{s}}lar, Ciobanu, Ciolfi, Cipriano, Cirone, Clara, Clark,
  Clearwater, Clesse, Cleva, Coccia, Cohadon, Cohen, Colleoni, Collette,
  Collins, Colpi, Constancio, Conti, Cooper, Corban, Corbitt,
  Cordero-Carri{\'o}n, Corezzi, Corley, Cornish, Corre, Corsi, Cortese, Costa,
  Cotesta, Coughlin, Coughlin, Coulon, Countryman, Couvares, Covas, Coward,
  Cowart, Coyne, Coyne, Creighton, Creighton, Cripe, Croquette, Crowder,
  Cudell, Cullen, Cumming, Cummings, Cunningham, Cuoco, Curylo, Canton,
  D{\'a}lya, Dana, Daneshgaran-Bajastani, D'Angelo, Danilishin, D'Antonio,
  Danzmann, Darsow-Fromm, Dasgupta, Datrier, Dattilo, Dave, Davier, Davies,
  Davis, Daw, DeBra, Deenadayalan, Degallaix, Laurentis, Del{\'e}glise,
  Delfavero, Lillo, Pozzo, DeMarchi, D'Emilio, Demos, Dent, Pietri, Rosa,
  Rossi, DeSalvo, Varona, \& Dhurand...}]{Abbott:2020jo}
Abbott, R., Abbott, T.~D., Abraham, S., {et~al.} 2020, The Astrophysical
  Journal Letters, 896, L44

\bibitem[{Abbott et~al . {et~al.}(2018)Abbott et~al ., Collaboration, Abbott,
  Abbott, Abbott, Abraham, Acernese, Ackley, Adams, Adhikari, Adya, Affeldt,
  Agathos, Agatsuma, Aggarwal, Aguiar, Aiello, Ain, Ajith, Allen, Allocca,
  Aloy, Altin, Amato, Ananyeva, Anderson, Anderson, Angelova, Antier, Appert,
  Arai, Araya, Areeda, Ar{\`e}ne, Arnaud, Arun, Ascenzi, Ashton, Aston, Astone,
  Aubin, Aufmuth, AultONeal, Austin, Avendano, Avila-Alvarez, Babak, Bacon,
  Badaracco, Bader, Bae, Baker, Baldaccini, Ballardin, Ballmer, Banagiri,
  Barayoga, Barclay, Barish, Barker, Barkett, Barnum, Barone, Barr, Barsotti,
  Barsuglia, Barta, Bartlett, Bartos, Bassiri, Basti, Bawaj, Bayley, Bazzan,
  B{\'e}csy, Bejger, Belahcene, Bell, Beniwal, Berger, Bergmann, Bernuzzi,
  Bero, Berry, Bersanetti, Bertolini, Betzwieser, Bhandare, Bidler, Bilenko,
  Bilgili, Billingsley, Birch, Birney, Birnholtz, Biscans, Biscoveanu, Bisht,
  Bitossi, \& Bizouard}]{Abbottetal:2018vb}
Abbott et~al ., B.~P., Collaboration, t.~V., Abbott, B.~P., {et~al.} 2018,
  1811.12940

\bibitem[{Antonini {et~al.}(2017)Antonini, Rodriguez, Petrovich, \&
  Fischer}]{Antonini:2017el}
Antonini, F., Rodriguez, C.~L., Petrovich, C., \& Fischer, C.~L. 2017, Monthly
  Notices of the Royal Astronomical Society: Letters, L58

\bibitem[{Bavera {et~al.}(2019)Bavera, Fragos, Qin, Zapartas, Neijssel, Mandel,
  Batta, Gaebel, Kimball, \& Stevenson}]{Bavera:2019ut}
Bavera, S.~S., Fragos, T., Qin, Y., {et~al.} 2019, 1906.12257

\bibitem[{Belczynski {et~al.}(2002)Belczynski, Kalogera, \&
  Bulik}]{Belczynski:2002gi}
Belczynski, K., Kalogera, V., \& Bulik, T. 2002, The Astrophysical Journal,
  572, 407

\bibitem[{Chatterjee {et~al.}(2016)Chatterjee, Rodriguez, Kalogera, \&
  Rasio}]{Chatterjee:2016fl}
Chatterjee, S., Rodriguez, C.~L., Kalogera, V., \& Rasio, F.~A. 2016, The
  Astrophysical Journal, L26

\bibitem[{Dominik {et~al.}(2012)Dominik, Belczynski, Fryer, Holz, Berti, Bulik,
  Mandel, \& O'Shaughnessy}]{Dominik:2012cwa}
Dominik, M., Belczynski, K., Fryer, C., {et~al.} 2012, The Astrophysical
  Journal, 759, 52

\bibitem[{D'Orazio {et~al.}(2016)D'Orazio, Haiman, Duffell, MacFadyen, \&
  Farris}]{DOrazio:2016gy}
D'Orazio, D.~J., Haiman, Z., Duffell, P., MacFadyen, A., \& Farris, B. 2016,
  Monthly Notices of the Royal Astronomical Society, 459, 2379

\bibitem[{Duffell {et~al.}(2019)Duffell, D'Orazio, Derdzinski, Haiman,
  MacFadyen, Rosen, \& Zrake}]{Duffell:2019vu}
Duffell, P.~C., D'Orazio, D., Derdzinski, A., {et~al.} 2019, arXiv.org,
  arXiv:1911.05506

\bibitem[{Ertl {et~al.}(2020)Ertl, Woosley, Sukhbold, \& Janka}]{Ertl:2020ga}
Ertl, T., Woosley, S.~E., Sukhbold, T., \& Janka, H.~T. 2020, The Astrophysical
  Journal, 890, 51

\bibitem[{Farr {et~al.}(2011)Farr, Sravan, Cantrell, Kreidberg, Bailyn, Mandel,
  \& Kalogera}]{Farr:2011ct}
Farr, W.~M., Sravan, N., Cantrell, A., {et~al.} 2011, The Astrophysical
  Journal, 741, 103

\bibitem[{Flannery \& van~den Heuvel(1975)}]{Flannery:1975to}
Flannery, B.~P., \& van~den Heuvel, E. P.~J. 1975, Astronomy {\&} Astrophysics,
  39, 61

\bibitem[{Fragione {et~al.}(2020)Fragione, Loeb, \& Rasio}]{Fragione:2020dc}
Fragione, G., Loeb, A., \& Rasio, F.~A. 2020, arXiv.org, L15

\bibitem[{Fryer {et~al.}(2012)Fryer, Belczynski, Wiktorowicz, Dominik,
  Kalogera, \& Holz}]{Fryer:2012jk}
Fryer, C.~L., Belczynski, K., Wiktorowicz, G., {et~al.} 2012, The Astrophysical
  Journal, 749, 91

\bibitem[{Gerosa {et~al.}(2018)Gerosa, Berti, O'Shaughnessy, Belczynski,
  Kesden, Wysocki, \& Gladysz}]{Gerosa:2018hw}
Gerosa, D., Berti, E., O'Shaughnessy, R., {et~al.} 2018, Physical Review D, 126

\bibitem[{Gupta {et~al.}(2020)Gupta, Gerosa, Arun, Berti, Farr, \&
  Sathyaprakash}]{Gupta:2020kz}
Gupta, A., Gerosa, D., Arun, K.~G., {et~al.} 2020, Physical Review D, 101,
  103036

\bibitem[{Jani \& Loeb(2020)}]{Jani:2020hl}
Jani, K., \& Loeb, A. 2020, The Astrophysical Journal Letters, 889, L35

\bibitem[{Kaaret {et~al.}(2017)Kaaret, Feng, \& Roberts}]{Kaaret:2017eb}
Kaaret, P., Feng, H., \& Roberts, T.~P. 2017, Annual Review of Astronomy and
  Astrophysics, 55, 303

\bibitem[{Kratter \& Lodato(2016)}]{Kratter:2016dw}
Kratter, K., \& Lodato, G. 2016, Annual Review of Astronomy and Astrophysics,
  54, 271

\bibitem[{Most {et~al.}(2020)Most, Papenfort, Weih, \& Rezzolla}]{Most:2020uu}
Most, E.~R., Papenfort, L.~J., Weih, L.~R., \& Rezzolla, L. 2020, arXiv.org,
  arXiv:2006.14601

\bibitem[{{\"O}zel {et~al.}(2010){\"O}zel, Psaltis, Narayan, \&
  McClintock}]{Ozel:2010hd}
{\"O}zel, F., Psaltis, D., Narayan, R., \& McClintock, J.~E. 2010, The
  Astrophysical Journal, 1918

\bibitem[{{Peters, P. C.}(1964)}]{PetersPC:1964bc}
{Peters, P. C.} 1964, Physical Review, 136, 1224

\bibitem[{{Portegies Zwart} {et~al.}(2004){Portegies Zwart}, {Baumgardt},
  {Hut}, {Makino}, \& {McMillan}}]{Zwart:2004jj}
{Portegies Zwart}, S.~F., {Baumgardt}, H., {Hut}, P., {Makino}, J., \&
  {McMillan}, S. L.~W. 2004, \nat, 428, 724

\bibitem[{Rimoldi {et~al.}(2015)Rimoldi, Zwart, \& Rossi}]{Rimoldi:2015up}
Rimoldi, A., Zwart, S.~P., \& Rossi, E.~M. 2015, arXiv.org, 1510.02483v1

\bibitem[{Rodriguez {et~al.}(2018)Rodriguez, Amaro-Seoane, Chatterjee, Kremer,
  Rasio, Samsing, Ye, \& Zevin}]{Rodriguez:2018ci}
Rodriguez, C.~L., Amaro-Seoane, P., Chatterjee, S., {et~al.} 2018, Physical
  Review D, 1

\bibitem[{Safarzadeh {et~al.}(2020)Safarzadeh, Hamers, Loeb, \&
  Berger}]{Safarzadeh:2020gs}
Safarzadeh, M., Hamers, A.~S., Loeb, A., \& Berger, E. 2020, The Astrophysical
  Journal Letters, 888, L3

\bibitem[{Schr{\o}der {et~al.}(2018)Schr{\o}der, Batta, \&
  Ramirez-Ruiz}]{Schroder:2018ko}
Schr{\o}der, S.~L., Batta, A., \& Ramirez-Ruiz, E. 2018, arXiv.org, L3

\bibitem[{Sukhbold {et~al.}(2018)Sukhbold, Woosley, \& Heger}]{Sukhbold:2018br}
Sukhbold, T., Woosley, S.~E., \& Heger, A. 2018, The Astrophysical Journal,
  860, 93

\bibitem[{Suwa {et~al.}(2015)Suwa, Yoshida, Shibata, Umeda, \&
  Takahashi}]{Suwa:2015bs}
Suwa, Y., Yoshida, T., Shibata, M., Umeda, H., \& Takahashi, K. 2015, Monthly
  Notices of the Royal Astronomical Society, 454, 3073

\bibitem[{Suzuki \& Maeda(2016)}]{Suzuki:2016fz}
Suzuki, A., \& Maeda, K. 2016, arXiv.org, 2633

\bibitem[{Tagawa {et~al.}(2019)Tagawa, Haiman, \& Kocsis}]{Tagawa:2019tu}
Tagawa, H., Haiman, Z., \& Kocsis, B. 2019, arXiv.org, arXiv:1912.08218

\bibitem[{Toomre(1964)}]{Toomre:1964fe}
Toomre, A. 1964, Astrophysical Journal, 139, 1217

\bibitem[{Truelove \& McKee(1999)}]{Truelove:1999it}
Truelove, J.~K., \& McKee, C.~F. 1999, The Astrophysical Journal Supplement
  Series, 120, 299

\bibitem[{Vattis {et~al.}(2020)Vattis, Goldstein, \&
  Koushiappas}]{Vattis:2020ub}
Vattis, K., Goldstein, I.~S., \& Koushiappas, S.~M. 2020, arXiv.org,
  arXiv:2006.15675

\bibitem[{{Yang} {et~al.}(2020){Yang}, {Gayathri}, {Bartos}, {Haiman},
  {Safarzadeh}, \& {Tagawa}}]{Yang2020}
{Yang}, Y., {Gayathri}, V., {Bartos}, I., {et~al.} 2020, arXiv e-prints,
  arXiv:2007.04781

\bibitem[{Zaldarriaga {et~al.}(2017)Zaldarriaga, Kushnir, \&
  Kollmeier}]{Zaldarriaga:2017fn}
Zaldarriaga, M., Kushnir, D., \& Kollmeier, J.~A. 2017, Monthly Notices of the
  Royal Astronomical Society, 473, 4174

\bibitem[{Zevin {et~al.}(2020)Zevin, Spera, Berry, \& Kalogera}]{Zevin:2020vp}
Zevin, M., Spera, M., Berry, C. P.~L., \& Kalogera, V. 2020, arXiv.org,
  2006.14573v1

\end{thebibliography}
\end{document}